\documentclass[11pt]{article} 
\usepackage{graphicx,floatflt,amssymb,epsf,rotate} 
\begin{document} 
 
\vskip 30pt 
 
\begin{center} 
{\Large \bf Dimension-5
operators and the }
\vskip 5pt 
{\Large \bf unification condition in SO(10) and E(6)} \\
\vspace*{1cm} 
\renewcommand{\thefootnote}{\fnsymbol{footnote}} 
{ {\sf Joydeep
Chakrabortty${}^1$\footnote{e-mail: joydeep@hri.res.in}  and}
 {\sf Amitava Raychaudhuri${}^{1,2}$} 
} \\ 
\vspace{10pt} 
   ${}^{1)}$ {\em Harish-Chandra Research Institute,\\
Chhatnag Road, Jhunsi, Allahabad  211 019, India}  

   ${}^{2)}$ {\em Department of Physics, University of Calcutta, \\
92 Acharya Prafulla Chandra Road, Kolkata 700 009, India}  \\
\normalsize 
\end{center} 
 
\begin{abstract} 
\noindent
Effective dimension-5 operators which modify the gauge kinetic
term in Grand Unified Theories may arise as a consequence of
quantum gravity or string compactification.  We exhaustively
calculate the modification of the gauge unification condition due
to such operators for all viable rank-preserving symmetry
breakings of SO(10) and E(6) grand unified models.

\vskip 5pt 
\noindent 
\texttt{Key Words:~~Grand Unified Theories } 
\end{abstract}

\renewcommand{\thesection}{\Roman{section}} 
\setcounter{footnote}{0} 
\renewcommand{\thefootnote}{\arabic{footnote}} 
 
\section{Introduction} 

The Standard Model (SM) is based on the gauge symmetry ${\mathcal
G}_{SM} \equiv SU(3)_c \otimes SU(2)_L \otimes U(1)_Y$ which has
three independent couplings $g_3, g_2,$ and $g_1$ of different
strength. If the strong and electroweak interactions merge at
high energy within a grand unified theory (GUT) framework then at
that stage they will be described by a single gauge coupling
$g_{GUT}$. This will necessitate a correlation among the
strengths of the three forces measured at lower energies ($\sim
M_{Z}$).  GUTs would also interrelate fermion masses 
because quark-lepton unification is an essential
ingredient of the programme \cite{guts, rg}.

The current low energy measured values of the gauge couplings are
no longer consistent with unification in a minimal $SU(5)$ GUTs
nor is the key prediction of proton decay so far observed.
Further, in $SU(5)$, the SM  Higgs doublet ($H_{2}$) sits in a
GUT multiplet with a colour triplet  ($H_{3}$).  While $H_{2}$
should be at the electroweak scale, $H_{3}$ needs to be very
heavy to avoid rapid proton decay  -- the so-called {\em
doublet-triplet splitting}.  Another shortcoming of the minimal
GUT concerns the fermion mass relations. Though $b-\tau$
unification occurs naturally, the light quark-lepton mass ratios,
such as $m_{s}/m_{\mu}$, do not follow.  This has encouraged
interest in GUT models based on larger groups such as $SO(10)$
and $E(6)$ which provide the option of a higher unification scale
suppressing proton decay along with a richer Higgs structure and
several intermediate symmetry-breaking mass scales leading to
testable consequences at colliders or {\em via} the observation
of $n-\bar{n}$ oscillations. Needless to say, unification of all
interactions with gravity is the final objective and grand
unification is a step in this direction.

In the absence of a full quantum theory of gravity it has been a
useful exercise to explore some of its implications on grand
unification through higher dimension gauge invariant effective
operators, suppressed by powers of the Planck mass, $M_{Pl}$.  In
string theory, similar effective operators could arise from
string compactification, $M_{Pl}$ being then replaced by the
compactification scale $M_c$.

In this work we focus on the corrections to the gauge kinetic
term:
\begin{equation}
\mathcal{L}_{kin}=-\frac{1}{4 c} Tr(F_{\mu\nu} F^{\mu\nu}) .
\label{eq:kin}
\end{equation}
where $F^{\mu\nu} = \Sigma_i \lambda_{i}.F_{i}^{\mu\nu}$ is the gauge field
strength tensor with $\lambda_i$ the matrix representations
of the generators of  ${\mathcal G}_{GUT}$
normalized to
$Tr(\lambda_{i}\lambda_{j})=c~\delta_{ij}$. The $\lambda_{i}$ are
often chosen in the fundamental representation with $c = 1/2$. In
the following, other representations are sometimes found convenient.

The dimension-5 term  from quantum gravity
(or string compactification) examined here is \cite{prev}:
\begin{equation}
\mathcal{L}_{dim-5}=-\frac{\eta}{M_{Pl}}\left[\frac{1}{4
c}Tr(F_{\mu\nu}\Phi_{D} F^{\mu\nu})\right]
\label{eq:dim5op}
\end{equation}
where $\Phi_{D}$ stands for the $D$-component scalar multiplet
and $\eta$ parametrises the strength of this interaction.  A
gauge invariant of the form in eq.  (\ref{eq:dim5op}) is allowed
if $\Phi_D$ is in any representation included in the symmetric
product of two adjoint representations of the group. For example,
in $SO(10)$, $(45 \otimes 45)_{sym} = 1 \oplus 54 \oplus 210
\oplus 770 $ and $\Phi_D$ may be 54- or 210- or 770-dimensional.

When  $\Phi_{D}$ develops a vacuum expectation value ({\em vev})
$v_D$, which sets the scale of grand unification $M_X$ and drives
the symmetry breaking\footnote{Since $\Phi_D$ arises from the symmetric
product of two adjoint representations the symmetry breaking is
rank preserving.} ${\mathcal G}_{GUT} \rightarrow {\mathcal
G}_{1} \otimes {\mathcal G}_{2} \otimes \ldots {\mathcal G}_{n}$,
an effective gauge kinetic term is generated from eq.
(\ref{eq:dim5op}). Depending on the structure of the {\em vev}, this
additional contribution, in general, will be different for the
kinetic terms for the subgroups ${\mathcal G}_{1}, \ldots
{\mathcal G}_{n}$. After an appropriate scaling of the
gauge fields this results in  a  modification of the gauge coupling
unification condition to:
\begin{equation}
g_{1}^{2}(M_{X})(1+\epsilon\delta_1)=g_{2}^{2}(M_{X})(1+\epsilon\delta_2)
= \ldots = g_{n}^{2}(M_{X})(1+\epsilon\delta_n)  ,
\label{eq:modunif}
\end{equation}
wherein the $\delta_i, ~i=1,2,\ldots n$,  arise from eq.
(\ref{eq:dim5op}) and $\epsilon = \eta v_{D}/2M_{Pl} \sim {\cal
O} (M_X/M_{Pl})$.  Thus, the presence of the dimension-5 terms in
the Lagrangian modify the usual boundary conditions on gauge
couplings, namely, that they are expected to unify at $M_X$. The
$\delta_i$ were calculated for a set of phenomenologically
interesting breaking sequences for $SU(5)$, $SO(10)$, and $E(6)$
in our earlier work
\cite{cr1}.  It is not impossible that the modification in eq.
(\ref{eq:modunif}) will enable the unification programme to
succeed with the current low energy values of the coupling
constants.  This was also examined in the
context of the above GUT groups \cite{cr1, cr2}.  In
supersymmetric GUTs, the ratio of these $\delta_i$ determine the
non-universal gaugino mass ratios \cite{cr1, nonug1, nonug2} whose
detectability at high energy colliders has been investigated in
detail.

In  earlier works the factors $\delta_i$ arising from the dim-5
operators were obtained for the $SU(5)$, $SO(10)$, and $E(6)$ GUT
groups for some selected breaking patterns\footnote{The
$\delta_i$ presented here differ from our earlier results in the
magnitudes.  The relative values are unchanged.}
\cite{prev, cr1}.  For $SO(10)$ and $E(6)$ those chains which
admit a left-right symmetry were picked. In this note we complete
the exercise; we work out the unification conditions in the
presence of dim-5 operators for {\em all} phenomenologically
viable symmetry descents of the parent group, which could be
either $SO(10)$ or $E(6)$, so long as the first step in rank
preserving. Thus for $SO(10)$ the options for the first step of
symmetry breaking are $SU(5) \otimes U(1)_X$, $SU(4)_c \otimes
SU(2)_L \otimes SU(2)_R$ and $SU(3)_c \otimes SU(2)_L \otimes
U(1)_Y \otimes U(1)_X$. For $E(6)$ these could be any
one of $SU(3)_c \otimes SU(3)_L \otimes SU(3)_R$, $SU(2) \otimes
SU(6)$, $SO(10) \otimes U(1)_\eta$, $SU(5) \otimes U(1)_\xi
\otimes U(1)_\eta$, $SU(3)_c
\otimes SU(2)_L \otimes U(1)_Y \otimes U(1)_\xi \otimes
U(1)_\eta$.  This work provides a compendium of the results for
all options.  After noting the known consequences for $SU(5)$, we
present the results for the different alternatives possible with
$SO(10)$ and $E(6)$ in the succeeding sections. We end with our
conclusions. The detailed forms of the {\em vev}s are relegated
to an Appendix.

\section{SU(5) GUT}\label{s:su5}

The case of $SU(5)$ has been discussed earlier in the literature
\cite{prev, cr1}. For completeness we summarise the results here.
$\Phi_D$ can be in the 24-, 75- or 200-dimensional
representation of $SU(5)$ and the symmetry is broken to $SU(3)_c
\otimes SU(2)_L \otimes U(1)_Y$. 

The procedure to obtain these results \cite{cr1} is to express
$<\Phi_D>$ as a diagonal matrix of dimensionality of some
$SU(5)$ irreducible representation. From the ${\mathcal G}_{SM}$
structure of this representation, the $\delta_i$ can be read off.
We list the $<\Phi_D>$ for the various cases in the Appendix (see
sec. \ref{a:su5}).

The $\delta_i$ arising in the different cases are listed
in Table \ref{tab:su5}.
\begin{table}
\begin{center}
\begin{tabular}{|c|c|c|c|c|} \hline
$SU(5)$  Representations & $\delta_1$ & $\delta_2$ & $\delta_3$ & $N$  \\
\hline 
{\bf 24}  & 1 & 3 & -2 & 2/$\sqrt{15}$ \\
\hline
{\bf 75}  & 5  & -3 & -1 & 8/15$\sqrt{3}$ \\
\hline
{\bf 200}  &  10   &    2 & 1 & 1/35$\sqrt{21}$\\
\hline
\end {tabular}
\caption{Effective contributions (see eq. (\ref{eq:modunif})) to
gauge kinetic terms from different Higgs representations in eq.
(\ref{eq:dim5op}) for $SU(5)\rightarrow SU(3)_c\otimes
SU(2)_L\otimes U(1)_Y$. $N$ is an overall normalisation which has
been factored out from the $\delta_i$.}
\label{tab:su5}
\end {center}
\end{table}

\section{SO(10) GUT}

$SO(10)$ \cite{so10} is now the widely preferred model for grand
unification, offering the option of descending to ${\mathcal
G}_{SM}$ through a left-right symmetric route \cite{lrs} -- the
intermediate Pati-Salam ${\mathcal G}_{PS} \equiv$ $SU(4)_c
\otimes SU(2)_L\otimes SU(2)_R$ -- or {\em via} $SU(5) \otimes
U(1)_{X}$ or in one step to $SU(3)_c \otimes SU(2)_L \otimes
U(1)_Y \otimes U(1)_X$.   The effect of dimension-5 interactions
in the first case has been reported upon before
\cite{cr1} and here we will only recapitulate those results and lay
primary emphasis on the other options.

\subsection{SO(10) $\rightarrow$ SU(5)$\otimes$U(1)} \label{s:10_51}
Under $SU(5) \otimes U(1)_X$ the $SO(10)$ spinorial
representation decomposes\footnote{The correctly normalised
($Tr(\lambda_{i}\lambda_{j})=2 ~\delta_{ij}$) $U(1)_X$ charges
are obtained by multiplying the displayed quantum numbers by a
factor of $\frac{1}{2\sqrt{10}}$ .}  as follows: 16 $\equiv$
(1,-5) + ($\bar5$,3) +  (10,-1). The SM families belong to this
representation.  The particle assignments within the 16-plet can
be chosen in two distinct ways with different physics
consequences: (a) {\bf conventional $\bf SU(5)$}: $U(1)_{X}$
commutes with the SM, so the low scale hypercharge ($U(1)_{Y}$)
is the same as the $U(1)_{Y'}$ in $SU(5)$; e.g., for the
($\bar5$,3) multiplet $T_{Y'} \equiv \sqrt{\frac{3}{5}}
~diag(\frac{1}{3}, \frac{1}{3}, \frac{1}{3},
-\frac{1}{2},-\frac{1}{2})$.  The SM generators are entirely
within the $SU(5)$  and a singlet under it is uncharged.
Therefore, the (1,-5) submultiplet has to be identified with the
neutral member in the 16-plet, the $\nu_i^c$ ($i$ = 1,2,3).  The
other option is (b) {\bf flipped $\bf SU(5)$}: Here $U(1)_{Y'}$
and $U(1)_X$ combine to give $U(1)_{Y}$: $T_{Y}= -(2\sqrt{6}
~T_{X} + T_{Y'}$)/5 \cite{fsu5}.  The difference can be
illustrated using the ($\bar5$,3) multiplet. For it the $U(1)_{Y'}$
assignment is, as before, $T_{Y'} \equiv \sqrt{\frac{3}{5}}
~diag(\frac{1}{3}, \frac{1}{3},\frac{1}{3}, -\frac{1}{2},
-\frac{1}{2})$ while the {\em normalised} $U(1)_{X}$ is
$\frac{3}{2\sqrt{10}}$ so that $T_{Y} \equiv \sqrt{\frac{3}{5}}
~diag(-\frac{2}{3}, -\frac{2}{3},-\frac{2}{3}, -\frac{1}{2},
-\frac{1}{2})$. Thus, this submultiplet now contains
$(u^c_i, L_i$) rather than the usual $(d^c_i, L_i)$. The (1,-5)
state is $SU(3)_c\otimes SU(2)_L$ singlet but carries a non-zero
hypercharge, $Y = 1$. The only particle that satisfies this
requirement is $l_i^c$.

The complete particle assignments for the first generation in the
two options are:

(a) For conventional $SU(5)$
\begin{eqnarray}
{\mathbf{(1,
-5)}}~=~\nu^c_1~,~~{\mathbf{(\bar 5,
3)}}~=~(d^c_1, ~l_1)~,~ ~~{\mathbf{(10, -1)}}~=~(q_1, ~u^c_1, ~e^c_1) ~,
\label{su5}
\end{eqnarray}
and (b) for $flipped$ $SU(5)$: 
\begin{eqnarray}
{\mathbf{(1, -5)}}~=~e^c_1,~~{\mathbf{(\bar
5, 3)}}~=~(u^c_1, ~l_1)~,~ ~~{\mathbf{(10, -1)}}~=~(q_1, ~d^c_1, ~\nu^c_1)~,~
\label{splitsu5}
\end{eqnarray}
where $q$ and $l$ are respectively the left-handed quark and
lepton doublets, $u^c$, $d^c$, $e^c$, and $\nu^c$ are the $CP$
conjugated states corresponding to the right-handed up-type
quark, down-type quark, lepton, and neutrino, respectively.

In $SO(10)$ GUT, at the unification scale one has $g_{5}=g_{1}$. 
The presence of any dim-5 effective
interactions of the form of eq. (\ref{eq:dim5op}) will affect
this relation generating corrections as shown in eq.
(\ref{eq:modunif}) which in this case will involve two parameters
$\delta_5$ and $\delta_1$.

As noted earlier, $\Phi_D$ can be chosen only in the 54, 210, and
770-dimensional representations. Of these,  the 54 does not have
an $SU(5) \otimes U(1)$ singlet. So,  only the 210- and
770-dimensional cases need examination.

\begin{table}
\begin{center}
\begin{tabular}{|c|c|c|c|} \hline
$SO(10)$  Representations & $\delta_{5}$ & $\delta_{1}$ & $N$   \\
\hline 
{\bf 210}  &  -1  & 4 &  1/4$\sqrt{5}$ \\
\hline
{\bf 770}  &  1 &  16 & -1/24$\sqrt{5}$ \\
\hline
\end {tabular}
\caption{Effective contributions (see eq. (\ref{eq:modunif})) to
gauge kinetic terms from different Higgs representations in eq.
(\ref{eq:dim5op}) for $SO(10)\rightarrow SU(5)\otimes U(1)$. $N$
is an overall normalisation which has been factored out from the $\delta_i$.}
\label{tab:so10_51}
\end {center}
\end{table}

Using $(\overline{16} \otimes 16) = 1 \oplus  45 \oplus 210 $,
$<\Phi_{210}>$ can be expressed as a 16-dimensional traceless
diagonal matrix. The form of this {\em vev} for this symmetry
breaking is given in (\ref{vev210_1}). It yields
$\delta_{5}=-\frac{1}{4\sqrt{5}}$ and
$\delta_{1}=\frac{1}{\sqrt{5}}$.

In a similar fashion the {\em vev} of $\Phi_{770}$ can be written
as the 45$\times$45 diagonal traceless matrix in
(\ref{vev770_1}).  This results in $\delta_{5} =
-\frac{1}{24\sqrt{5}}$, $\delta_{1} = -\frac{2}{3\sqrt{5}}$.

The results for this chain of symmetry breaking are summarised in
Table \ref{tab:so10_51}. The $\delta_{i}$ are completely group
theoretic in nature and obviously do not depend on whether the
particle assignments follow the conventional or flipped $SU(5)$.

\subsection{SO(10) $\rightarrow$ SU(3)$_c \otimes$ SU(2)$_L
\otimes$ U(1)$_Y \otimes$ U(1)$_X$}\label{s:10_3211}

\begin{table}
\begin{center}
\begin{tabular}{|c|c|c|c|c|c|} \hline
$SO(10)$  Representations & $\delta_1$ & $\delta_2$ &
$\delta_3$&$\delta_{1X}$ & $N$  \\ \hline 
{\bf 54 (24)}  & 1 & 3 & -2 & 0  & 1/2$\sqrt{15}$\\
\hline 
{\bf 210 (24)}  & 1 & 3 & -2 & 0  & 1/4$\sqrt{15}$\\
{\bf 210 (75)}  &  5  & -3 & -1 & 0 & 1/12 \\
\hline 
{\bf 770 (24)} & 1 & 3 & -2 & 0  & 2/$\sqrt{15}$\\
{\bf 770 (75)}  &  5  & -3 & -1 & 0 & 8/15$\sqrt{3}$ \\
{\bf 770 (200)}  &  10  &    2 & 1 & 0  & -1/8$\sqrt{21}$\\ \hline
\end {tabular}
\caption{Effective contributions (see eq. (\ref{eq:modunif})) to
gauge kinetic terms from different Higgs representations in eq.
(\ref{eq:dim5op}) for $SO(10)\rightarrow SU(3)_c\otimes
SU(2)_L\otimes U(1)_Y\otimes U(1)_X$. $SU(5)$ subrepresentations
are indicated within parentheses. $N$ is an overall normalisation which has
been factored out from the $\delta_i$.}
\label{tab:so10_3211}
\end {center}
\end{table}

The unification condition in the presence of dimension-5 effective
interactions of the form of eq. (\ref{eq:dim5op}) will now
involve the parameters $\delta_i$ ($i$ = 1,2,3) as for $SU(5)$
and an additional one $\delta_{1X}$.

In order to break $SO(10)$ directly to ${\mathcal G}_{3211}
\equiv SU(3)_c \otimes SU(2)_L \otimes U(1)_Y \otimes U(1)_X$ the
{\em vev} must be a non-singlet not just under $SO(10)$ but also
under $SU(5)$. The decompositions of $SO(10)$ representations
under $SU(5)\otimes U(1)$ are useful for identifying these {\em
vev}s. The calculation can be considerably  simplified by using
the $SU(5)$ symmetry breaking patterns at our disposal from sec.
\ref{s:su5}. One simply has to look for 24, 75, and 200
submultiplets within the 54, 210, and 770 $SO(10)$ multiplets.

The 54 representation of $SO(10)$ has a singlet under ${\mathcal
G}_{3211}$ which is contained in a 24 of $SU(5)$. The {\em vev}
for this case is shown in (\ref{vev54_2}) and the contributions
to the $\delta_i$ can be immediately read off from the $SU(5)$
result in Table \ref{tab:su5}. These $\delta_i$ are listed in Table
\ref{tab:so10_3211}.

Notice, that in this case the effect of dimension-5 terms cannot
distinguish between an $SU(5)$ theory with $<\Phi_{24}>$ driving
the symmetry breaking and an $SO(10)$ one with $<\Phi_{54}>$.  For
$\Phi_{210}$ or $\Phi_{770}$ the situation is different as they
have multiple ${\mathcal G}_{3211}$ singlet directions.

$\Phi_{210}$ has three directions which are all singlets under
${\mathcal G}_{3211}$.  Of these one is also an $SU(5)$ singlet.
In the subspace defined by them, three convenient orthogonal
directions can be identified, all singlets under $U(1)_X$, and
corresponding to 1-, 24- and 75-directions of the $SU(5)$
subgroup. If the {\em vev} is along one of these
directions\footnote{For the $SU(5)$ singlet direction the
$\delta_i$ are all equal. A {\em vev} in this direction alone
will not break $SO(10)$ to ${\mathcal G}_{3211}$.} it can be
simply read off from the results of section \ref{s:su5}. The {\em
vev}s corresponding to the 24 and 75 directions are given in
(\ref{vev210_2a}) and (\ref{vev210_2b}).  The $\delta_i$ derived
therefrom are shown in Table \ref{tab:so10_3211}. In general, we
have $<\Phi_{210}> = \alpha_{1}<\Phi_{210,1}>
+\alpha_{24}<\Phi_{210,24}> +  \alpha_{75}<\Phi_{210,75}>$, where
the $\alpha_i$ are complex numbers and the concomitant $\delta_i$
will be appropriately weighted combinations of the above results.

$<\Phi_{770}>$ has four ${\mathcal G}_{3211}$ invariant directions
which can be classified under the $SU(5)$ representations 1, 24,
75, and 200.  The  results for these are also shown in Table
\ref{tab:so10_3211}. Here again, in general, the {\em vev} may
lie in an arbitrary direction in the space spanned by the four
$SU(5)$-identified ones and the resultant $\delta_i$ can be
readily obtained from the above.  

Unlike the case of $<\Phi_{54}>$ where the singlet direction is
unique, the  $<\Phi_{210}>$ and  $<\Phi_{770}>$ provide a more
general option and therefore the predictions for the $\delta_i$
are not unique but cover a range. In this sense the model becomes
less predictive\footnote{This also applies to the 
non-universality options for gaugino masses in supersymmetric theories.}.

This route of symmetry breaking of $SO(10)$ {\em does not} ~admit
the flipped $SU(5)$ option by itself since in that case the
$U(1)_X$ combines with a $U(1)$ subgroup of $SU(5)$ to produce
$U(1)_Y$ and thus $SO(10)$ is broken to  ${\mathcal G}_{SM}$,
which is of rank 4, not 5. So this symmetry breaking  will have to
be through some other $SO(10)$ scalar multiplet. Nonetheless,
assuming that such a symmetry breaking is operational, we may ask
what would be the impact of the {\em vev}s of $\Phi_{54}$,
$\Phi_{210}$, and $\Phi_{770}$ of this subsection on the
unification parameters $\delta_i, ~i =1,2,3$. Using the {\em
vev}s used before and noting that $U(1)_{Y}$: $T_{Y}= -(2\sqrt{6}
~T_{X} + T_{Y'}$)/5 one finds the results presented in Table
\ref{tab:so10_321}.

\begin{table}
\begin{center}
\begin{tabular}{|c|c|c|c|c|} \hline
$SO(10)$  Representations & $\delta_1$ & $\delta_2$ &
$\delta_3$ & $N$  \\ \hline 
{\bf 54 (24)}  & 1 & 3 & -2   & 1/2$\sqrt{15}$\\
\hline 
{\bf 210 (1)}  & -19/5 & 1 & 1   & -1/4$\sqrt{5}$\\
{\bf 210 (24)}  & -7/5 & 3 & -2  & 1/4$\sqrt{15}$\\
{\bf 210 (75)}  &  1/5  & -3 & -1 & 1/12 \\
\hline 
{\bf 770 (1)}  & 77/5 & 1 & 1   & -1/24$\sqrt{5}$\\
{\bf 770 (24)} & -71/25 & 3 & -2 & 2/$\sqrt{15}$\\
{\bf 770 (75)}  &  1/5  & -3 & -1  & 8/15$\sqrt{3}$ \\
{\bf 770 (200)}  &  2/5  &    2 & 1 & -1/8$\sqrt{21}$\\ \hline
\end {tabular}
\caption{Effective contributions (see eq. (\ref{eq:modunif})) to
gauge kinetic terms from different Higgs representations in eq.
(\ref{eq:dim5op}) for $SO(10)\rightarrow SU(3)_c\otimes
SU(2)_L\otimes U(1)_Y$ (flipped $SU(5)$). $SU(5)$ subrepresentations
are indicated within parentheses. $N$ is an overall normalisation which has
been factored out from the $\delta_i$.}
\label{tab:so10_321}
\end {center}
\end{table}


\subsection{SO(10) $\rightarrow$ SU(4)$_c \otimes$SU(2)$_L
\otimes$SU(2)$_R$}

The left-right symmetric route of descent of $SO(10)$ has been
examined by us earlier in detail \cite{cr1, cr2}. Here for the
sake of completeness we briefly recall the results for the
symmetry breaking $SO(10) \rightarrow {\mathcal G}_{PS}$. 
As noted in eq. (\ref{eq:modunif}), we will be interested in the
factors $\delta_{4c}$, $\delta_{2L}$, and $\delta_{2R}$.

As before, $\Phi_D$ can be chosen only in the 54-, 210-, and
770-dimensional representations ensuring that $<\Phi_D >$ leaves
${\mathcal G}_{PS}$ unbroken.

For $\Phi_{54}$ the appropriate {\em vev} is given in
(\ref{vev54_3}) and this results in  $\delta_{4c} =
-{1\over{\sqrt{15}}}$ and $\delta_{2L} = \delta_{2R} =
{3\over{2\sqrt{15}}}$.  Notice that this correction to
unification ensures that $g_{2L}(M_X) = g_{2R}(M_X)$, i.e.,
D-parity \cite{dpar} is preserved.
\begin{table}
\begin{center}
\begin{tabular}{|c|c|c|c|c|} \hline
$SO(10)$  Representations & $\delta_{4c}$ & $\delta_{2L}$ &
$\delta_{2R}$ & $N$   \\
\hline 
{\bf 54}  & -2 & 3 & 3 & 1/$2\sqrt{15}$  \\
\hline
{\bf 210}  &  0 & 1 & -1 & 1/$2 \sqrt2$\\
\hline
{\bf 770}  &   2 &  5   & 5 & 1/$24 \sqrt5$ \\
\hline
\end {tabular}
\caption{Effective contributions (see eq. (\ref{eq:modunif})) to
gauge kinetic terms from different Higgs representations in eq.
(\ref{eq:dim5op}) for $SO(10)\rightarrow SU(4)_c \otimes
SU(2)_L \otimes SU(2)_R$. $N$ is an overall normalisation which has
been factored out from the $\delta_i$.}
\label{tab:so10_422}
\end {center}
\end{table}

A 16$\times$16 form of  $<\Phi_{210}>$ is given in
(\ref{vev210_3}) from which one can calculate $\delta_{4c} = $0
and $\delta_{2L} = -\delta_{2R} = {1\over2\sqrt2}$. D-parity is
broken through $<\Phi_{210}>$ and thus $g_{2L}(M_X) \neq
g_{2R}(M_X)$ though $SU(2)_L \otimes SU(2)_R$ remains unbroken at
$M_X$.

The final option is $\Phi_{770}$. One can write the {\em vev} in
terms of a 45-dimensional diagonal traceless matrix and this is
given in (\ref{vev770_3}).  From this one finds  $\delta_{4c} =
{1\over{12\sqrt{5}}}$ and the D-parity conserving $\delta_{2L} =
\delta_{2R} = {5\over{24\sqrt{5}}}$.

The results for this chain of $SO(10)$ breaking are collected
together in Table \ref{tab:so10_422}.

\section{E(6) GUT}

The exceptional group $E(6)$ has been proposed as a viable GUT
symmetry \cite{e6}. Among its subgroups of same rank are
$SU(3)_c$ $\otimes$ $SU(3)_L$ $\otimes$ $SU(3)_R$,
$SU(3)_c\otimes SU(2)_L \otimes U(1)_L \otimes SU(2)_R \otimes
U(1)_R$,   $SU(2)
\otimes SU(6)$, $SO(10) \otimes U(1)_\eta$,
$SU(5)\otimes U(1)_\xi\otimes U(1)_\eta$, and $SU(3)_c\otimes
SU(2)_L \otimes U(1)_Y \otimes U(1)_\xi \otimes U(1)_\eta$. All of these
intermediate gauge groups accommodate ${\mathcal G}_{SM}$ as a
subgroup and lead to different low scale phenomenology. We have
examined the first option in detail earlier
\cite{cr1, cr2}.  Here we will concentrate on the remaining
breaking chains recalling the first only for the sake of
completeness.

For $E(6)$ the adjoint representation is 78-dimensional.
Noting that $(78 \otimes 78)_{sym} = 1 \oplus  650 \oplus 2430$
it is clear that $\Phi_D$ can be either 650- or 2430-dimensional.
Both of them contain singlets under the above mentioned
intermediate gauge groups we are interested in.

\subsection{E(6) $\rightarrow$ SU(2)$\otimes$SU(6)}
\begin{table}
\begin{center}
\begin{tabular}{|c|c|c|c|} \hline
$E(6)$  Representations & $\delta_{2}$ & $\delta_{6}$ & $N$ \\
\hline 
{\bf 650}  &  5 &  -1 &  1/6$\sqrt{5}$\\
\hline
{\bf 2430}  &  -35 & -9 & 1/12 $\sqrt{910}$ \\
\hline
\end {tabular}
\caption{Effective contributions  (see eq. (\ref{eq:modunif})) to
gauge kinetic terms from different Higgs representations in eq.
(\ref{eq:dim5op}) for $E(6) \rightarrow SU(2) \otimes SU(6)$. $N$
is an overall normalisation which has been factored out from the $\delta_i$.}
\label{tab:e6_6x2}
\end {center}
\end{table}

The inconsistency  of the Georgi-Glashow $SU(5)$ model with the
proton decay and gauge unification requirements has been a
motivation to seek alternative GUT models. $SU(6)$ is one of
them.  It can naturally guarantee strong-CP invariance and a
supersymmetrised version implements doublet-triplet splitting by
the missing {\em vev} mechanism
\cite{su6}.

The subgroups from the breaking $E(6) \rightarrow SU(2) \otimes
SU(6)$ have been identified in several physically distinct
manners: $SU(2)_R \otimes SU(6)'$, $SU(2)_L \otimes SU(6)''$, and
$SU(2)_X \otimes SU(6)$. The results that we discuss are valid
irrespective of these alternative interpretations.

The contributions from the 650-dimensional representation for
this symmetry breaking chain can be obtained from eq.
(\ref{vev650_1}). One finds $\delta_{2} = \frac{5}{6\sqrt{5}}$
and $\delta_{6} = -\frac{1}{6\sqrt{5}}$,

For the 2430-dimensional $E(6)$ representation the {\em vev} is
given in (\ref{vev2430_1}). From it we get $\delta_{2} =
-\frac{35}{12\sqrt{910}}$, $\delta_{6} = -\frac{9}{6\sqrt{910}}$.
The results for this symmetry breaking chain can be found in
Table \ref{tab:e6_6x2}.

\subsection{E(6) $\rightarrow$ SO(10)$\otimes$U(1)$_\eta$}

$E(6)$ contains $SO(10) \otimes U(1)$ as a maximal subgroup.
Breaking patterns based on this are well discussed in the literature
\cite{e6}. Here, we consider the effect of dim-5 operators on the
gauge unification condition.

$<\Phi_{650}>$ is given in (\ref{vev650_2}). From it one obtains
$\delta_{10} = -\frac{1}{6\sqrt{5}}$, $\delta_{1} =
\frac{5}{6\sqrt{5}}$. Using (\ref{vev2430_2}) for $<\Phi_{2430}>$
one can similarly get $\delta_{10} = \frac{1}{72\sqrt{26}} $,
$\delta_{1} =\frac{3}{8\sqrt{26}}$.  These results are listed in
Table \ref{tab:e610x1}.

\begin{table}[bth]
\begin{center}
\begin{tabular}{|c|c|c|c|} \hline
$E(6)$  Representations & $\delta_{10}$ & $\delta_{1}$ & $N$  \\
\hline 
{\bf 650}  &  -1 &  5 &  1/6$\sqrt{5}$ \\
\hline
{\bf 2430}  &  1 & 27 &  1/72$\sqrt{26} $\\
\hline
\end {tabular}
\caption{Effective contributions  (see eq. (\ref{eq:modunif})) to
gauge kinetic terms from different Higgs representations in eq.
(\ref{eq:dim5op}) for $E(6) \rightarrow SO(10) \otimes U(1)$. $N$
is an overall normalisation which has been factored out from the $\delta_i$.}
\label{tab:e610x1}
\end {center}
\end{table}

\subsection{E(6) $\rightarrow$ SU(5)$ \otimes$U(1)$_\xi
\otimes$U(1)$_\eta$}

The results in this case are very similar to that for sec.
\ref{s:10_51}. There it was noted that the $SO(10)$ 210 and 770
representations contain singlets under $SU(5) \otimes U(1)_X$ and
the $\delta_5$ and $\delta_1$ in the two cases were presented in
Table \ref{tab:so10_51}. These results can be immediately taken
over for the current case with the change that the $U(1)_X$ is
here termed $U(1)_\xi$ and that $\delta_\eta = 0$ in all cases. 

The two relevant representations of $E(6)$ are 650 and 2430.
Of these, 650 contains a (210,0) submultiplet under
$SO(10)\otimes U(1)_\eta$. So for the 650 the $\delta_i$ will be
exactly as for the 210 in Table \ref{tab:so10_51}.

The  $E(6)$ 2430 representation contains both the (210,0) as well
as the (770,0) within it. If the {\em vev} is assigned to any one of
these directions the resultant $\delta_i$ will be as in the
respective case in Table \ref{tab:so10_51}. In general, the {\em
vev} will be a superposition of these two and so the $\delta_i$
will be the appropriately weighted value.

\subsection{E(6) $\rightarrow$ SU(3)$_c \otimes$SU(2)$_L
\otimes$U(1)$_Y \otimes$ U(1)$_\xi \otimes$U(1)$_\eta$}

As for the previous subsection, this alternative can be disposed
off straightforwardly using the results of  sec.
\ref{s:10_3211}. This time there is one extra step. In sec.
\ref{s:10_3211} results are presented for the $SO(10)$
representations 54, 210, and 770.  They can be immediately taken
over by noting that the $E(6)$ 650 contains (54,0) and (210,0)
submultiplets while the 2430 contains (54,0), (210,0), and (770,0).

The main changes compared to sec. \ref{s:10_3211} are that the
$U(1)_X$ there is called $U(1)_\xi$ here and for all cases
$\delta_\xi = \delta_\eta = 0$. For the 650 representation if the
{\em vev} is chosen along either the (54,0) or the (210,0)
directions then the results of Table \ref{tab:so10_3211} apply
directly. In general, of course, the $\delta_i$ will be a
weighted combination of these. Similar conclusions can be drawn
about the $<\Phi_{2430}>$ except that here, in general, the
$\delta_i$ will be a linear combination of the ones in 
 Table \ref{tab:so10_3211}.

\subsection{E(6) $\rightarrow$ SU(3)$_c \otimes$SU(3)$_L
\otimes$SU(3)$_R$}
This breaking chain \cite{e6_333} is exhaustively considered in
\cite{cr1,cr2} in the light of left-right symmetry \cite{lrs}.  A
$Z_2$ symmetry -- D-Parity -- is assumed to be active between
$SU(2)_L$ and $SU(2)_R$. The {\em vev} $<\Phi>$  can be
classified by its D-Parity behaviour.  $<\Phi_{650}>$ has two
directions which are singlets under ${\mathcal G}_{333}$ which
are even and odd under D-Parity.

The form of $<\Phi_{650}>$ is given in (\ref{vev650_5e}) for the
D-Parity even case while (\ref{vev650_5o}) is for the D-parity odd
{\em vev}.  This results in $\delta_{3c} = -{1\over{3\sqrt{2}}}$
and $\delta_{3L} = \delta_{3R} = {1\over{6\sqrt{2}}}$ for the
former and $\delta_{3c} = $0 and $\delta_{3L} = -\delta_{3R} =
{1\over{2\sqrt{6}}}$ for the latter.  $<\Phi_{2430}>$ is listed
in (\ref{vev2430_5}).  From it one can readily read off
$\delta_{3c} = \delta_{3L} = \delta_{3R} = -{1\over{4\sqrt{26}}}$.
In Table \ref{tab:e6_333} we collect the findings for the different
representations of $E(6)$.

It is worth remarking that the three $SU(3)$ subgroups in this
chain  are on an equal footing. It is possible to relate {\em
any} two of them through a $Z_2$-type discrete symmetry.  For the
purpose of illustration and for phenomenological interest we have
identified it with D-Parity. Obviously, one could just
as well choose the $Z_2$-type symmetry to be between $SU(3)_c$
and $SU(3)_L$ (or $SU(3)_R$).  The symmetry breaking {\em vev}s
of $\Phi_{650}$,
either even or odd under this changed parity-like symmetry, are
simply linear combinations between the {\em vev}s which are odd
and even under D-Parity discussed above.

\begin{table}
\begin{center}
\begin{tabular}{|c|c|c|c|c|} \hline
$E(6)$  Representations & $\delta_{3c}$ & $\delta_{3L}$ &
$\delta_{3R}$ & $N$   \\
\hline 
{\bf 650}  & -2 &  1 &  1 & -1/6$\sqrt2$ \\
\hline
{\bf 650$^\prime$}  & 0 &  1 &  -1&  1/2$\sqrt6$ \\
\hline
{\bf 2430}  &  1 & 1 & 1 &  -${1/4{\sqrt{26}}}$\\
\hline
\end {tabular}
\caption{Effective contributions  (see eq. (\ref{eq:modunif})) to
gauge kinetic terms from different Higgs representations in eq.
(\ref{eq:dim5op}) for $E(6) \rightarrow SU(3)_c \otimes SU(3)_L
\otimes SU(3)_R$. Note that there are two $SU(3)_c \otimes
SU(3)_L \otimes SU(3)_R$ singlet directions in 650. $N$ is an
overall normalisation which has been factored out from the $\delta_i$.}
\label{tab:e6_333}
\end {center}
\end{table}

\section{Conclusions}
Higher dimensional non-renormalisable interactions can partially
mimic the effects arising from quantum gravity or string
compactification. We have considered a class of such operators,
see eq. (\ref{eq:dim5op}), those that modify the gauge kinetic
term in GUTs. An operator of this type  involves a scalar
multiplet which, it turns out, can only break the symmetry in a
rank preserving manner. After spontaneous symmetry breaking such
terms affect the unification of coupling constants in a
calculable manner. The modifications depend on the subgroup to
which the GUT is broken at the first stage and are quantified in
terms of group theoretic factors $\delta_i$.  In supersymmetric
gauge theories the same factors are of interest as they
characterise non-universality of gaugino masses at the GUT scale.

For $SO(10)$ and $E(6)$ GUTs we have obtained the $\delta_i$ for
all allowed operators and all rank-preserving symmetry breakings.
For $SU(5)$ the symmetry breaking is unique. For some symmetry
breaking chains of $SO(10)$ (e.g., $SO(10) \rightarrow SU(5)
\otimes U(1)$) and $E(6)$, (e.g., $E(6) \rightarrow SO(10)
\otimes U(1)$) there is exactly one direction in a scalar
multiplet to which the {\em vev} can be ascribed and here again
the predictions are one-to-one. This is not so for some other
possibilities, (e.g., $SO(10) \rightarrow SU(3)_c \otimes SU(2)_L
\otimes U(1)_Y \otimes U(1)_X$) where multiple directions within a
scalar multiplet can accomplish the desired symmetry breaking. Here,
the predictions for gauge coupling unification are more flexible,
as are the implications for gaugino mass non-universality.

\vskip 20pt

{\bf Acknowledgements} This research has been supported by funds
from the XIth Plan `Neutrino physics' and RECAPP projects at HRI.

\setcounter{section}{0} 
\renewcommand{\thesection}{\Alph{section}}

\setcounter{equation}{0} 
\renewcommand{\theequation}{\Alph{section}.\arabic{equation}}

\section{Appendix: The vacuum expectation values}
In this Appendix we collect the different vacuum
expectation values which are used in this work. 

\subsection{SU(5)}\label{a:su5}
For $SU(5)$ $\Phi_D$ can be in the 24, 75, and 200 representations.

The prototype example of the vacuum expectation values found
useful in the calculations is in the case of $SU(5)$ with a
$\Phi_{24}$ scalar.  The {\em vev} of this field can be
represented as a traceless 5$\times$5 diagonal matrix
($Tr(\lambda_{i}\lambda_{j})=1/2 ~\delta_{ij}$):
\begin{equation}
<\Phi_{24}>=v_{24}\frac{1}{\sqrt{60}}~diag(3,3,-2,-2,-2) \equiv
v_{24} <24>_{5}.
\label{vev24_1}
\end{equation}
In addition 10- and 15-dimensional forms of the {\em vev},
identified through the property that the resulting $\delta_i$ are
the same as from (\ref{vev24_1}), are
also found useful. Under  ${\mathcal G}_{SM}$ the $SU(5)$ 10 =
 (1,1)$_{2}$ + ($\bar{3}$,1)$_{-{4\over3}}$ + (3,2)$_{1\over3}$.
Noting that
$Tr(\lambda_{i}\lambda_{j})=3/2 ~\delta_{ij}$, one finds:
\begin{equation}
<\Phi_{24}>=v'_{24}\frac{1}{\sqrt{60}}~diag(6,-4,-4,-4,
\underbrace{1,\ldots,1}_{6 ~entries}) \equiv v'_{24} <24>_{10}.
\label{vev24_2}
\end{equation}
Under ${\mathcal G}_{SM}$ the 15 of $SU(5)$
is (6,1)$_{-{4\over3}}$ + (3,2)$_{1\over3}$ + (1,3)$_{2}$ and 
one has ($Tr(\lambda_{i}\lambda_{j})=7/2 ~\delta_{ij}$):
\begin{equation}
<\Phi_{24}>=v''_{24}\frac{1}{\sqrt{60}}~diag(\underbrace{-4,\ldots,-4}_{6
~entries},\underbrace{1,\ldots,1}_{6
~entries},6,6,6) \equiv v''_{24} <24>_{15}.
\label{vev24_3}
\end{equation}
(\ref{vev24_1}), (\ref{vev24_2}), and (\ref{vev24_3}) 
yield the same $\delta_i$ if $v_{24} = v'_{24} = 9 v''_{24}$.

It comes of use for the discussions of $SO(10)$ to also list the
24$\times$24 forms of the different $SU(5)$ {\em vev}s. In this
case $Tr(\lambda_{i}\lambda_{j})=5 ~\delta_{ij}$. The 24 of
$SU(5)$ is (1,1)$_{0}$ + (1,3)$_{0}$ + (8,1)$_{0}$ +
(3,2)$_{-{5\over3}}$ + ($\bar{3}$,2)$_{5\over3}$. Thus
\begin{equation}
<\Phi_{24}>=v'''_{24}\sqrt{\frac{5}{252}}~diag(2,6,6,6,
\underbrace{-4,\ldots,-4}_{8 ~entries}, \underbrace{1,\ldots,1}_{6~entries},
\underbrace{1,\ldots,1}_{6~entries})  \equiv v'''_{24} <24>_{24}.
\label{vev24_4}
\end{equation}

For the {\em vev} of the 75-dimensional representation one uses
the $SU(5)$ relation: $10 \otimes \overline{10} = 1 \oplus 24
\oplus 75$. Taking into consideration that $<\Phi_{75}>$ must be
orthogonal to $<\Phi_{24}>$, i.e.,  (\ref{vev24_2}), it can be
expressed as:
\begin{equation}
<\Phi_{75}>=v_{75}\frac{1}{\sqrt{12}}~diag(3,1,1,1,
\underbrace{-1,\ldots,-1}_{6~entries}) \equiv v_{75} <75>_{10}.
\label{vev75_1}
\end{equation}
The 24$\times$24 form of $<\Phi_{75}>$ which yields the same
$\delta_i$ as (\ref{vev75_1}) is:
\begin{equation}
<\Phi_{75}>=v'_{75}\sqrt{\frac{5}{72}}~diag(5,-3,-3,-3,
\underbrace{-1,\ldots,-1}_{8 ~entries},
\underbrace{1,\ldots,1}_{6~entries},\underbrace{1,\ldots,1}_{6~entries})
\equiv v'_{75} <75>_{24}.
\label{vev75_2}
\end{equation}

Similarly, the relation $15 \otimes \overline{15} = 1 \oplus 24
\oplus 200$ permits the $vev$ for $\Phi_{200}$ to be written as a
15$\times$15 traceless diagonal matrix.  Ensuring
orthogonality with $<\Phi_{24}>$, i.e., (\ref{vev24_3}), one has:
\begin{equation}
<\Phi_{200}>=v_{200}\frac{1}{\sqrt{12}}~diag
(\underbrace{1,\ldots,1}_{6~entries},
\underbrace{-2,\ldots,-2}_{6~entries},2,2,2) \equiv v_{200} <200>_{15}.
\label{vev200_1}
\end{equation}
$<\Phi_{200}>$ can be also cast in a 24$\times$24 form. Keeping
(\ref{vev24_4}), (\ref{vev75_2}), and (\ref{vev200_1}) in mind, it is
found to be:  
\begin{equation}
<\Phi_{200}>=v'_{200}\sqrt{\frac{5}{168}}~diag(10,2,2,2,
\underbrace{1,\ldots,1}_{8 ~entries},\underbrace{-2,\ldots,-2}_{6~entries},
\underbrace{-2,\ldots,-2}_{6~entries}) \equiv v'_{200} <200>_{24}.
\label{vev200_2}
\end{equation}

\subsection{SO(10)}

For $SO(10)$ the possible choices for $\Phi_D$ are the 54-, 210-, and
770-dimensional representations.

The $SO(10)$ relation $10 \otimes 10 = 1 \oplus  45 \oplus 54$
ensures that $<\Phi_{54}>$ can be expressed as a
10$\times$10 traceless diagonal matrix. It is readily checked
that the normalisation condition is $Tr(\lambda_{i}\lambda_{j})=
~\delta_{ij}$. 

Similarly, $\overline{16} \otimes 16 = 1 \oplus  45 \oplus 210 $
permits $<\Phi_{210}>$ to be represented in a 16$\times$16
traceless diagonal form. For the 16$\times$16 matrices
$Tr(\lambda_{i}\lambda_{j})=2 ~\delta_{ij}$.

Finally, $<\Phi_{770}>$ can be written as a 45$\times$45 matrix
which is traceless and diagonal since $(45 \otimes 45)_{sym} = 1
\oplus 54 \oplus 210 \oplus 770$. Note that $<\Phi_{54}>$ and
$<\Phi_{210}>$ can also be written in a similar form and
orthogonality with them has to be ensured when obtaining
$<\Phi_{770}>$. For these matrices $Tr(\lambda_{i}\lambda_{j})=8
~\delta_{ij}$.

The above observations for $SO(10)$ are valid no matter which
chain of symmetry breaking is under consideration. 

\subsubsection{SO(10) $\rightarrow$ SU(5)$\otimes$U(1)}\label{ss:10_1}

For $\Phi_{54}$ there is no $SU(5)\otimes U(1)_X$ invariant direction.

Under $SU(5)\otimes U(1)_X$, 16 = (1,-5) + ($\bar 5$,3) +
(10,1).  Further, the diagonal matrix $<\Phi_{210}>$ must be
orthogonal to the one corresponding to $U(1)_X$, i.e.,
$\frac{1}{2\sqrt{10}}~diag$(-5,3,3,3,3,3,-1,-1,-1,-1,-1,-1,-1,-1,-1,-1).
Satisfying these, we find:
\begin{equation}
<\Phi_{210}>=v_{210}\frac{1}{\sqrt{20}}~diag(5,1,1,1,1,1,
\underbrace{-1,\ldots,-1}_{10 ~entries},)\equiv v_{210} <210>_{16}.
\label{vev210_1}
\end{equation}
Under $SU(5)$ $\otimes$ $U(1)_X$
45 = (1, 0) + (10, 4) + ($\overline{10}$, -4) + (24, 0). Asking the
results from (\ref{vev210_1}) be reproduced one arrives at:
\begin{equation}
<\Phi_{210}>=v'_{210}\sqrt{\frac{2}{15}}~diag(-4,\underbrace{-1,\ldots,-1}_{10
~entries},\underbrace{-1,\ldots,-1}_{10
~entries},\underbrace{1,\ldots,1}_{24 ~entries})\equiv v_{210} <210>_{45}.
\label{vev210_1a}
\end{equation}
$<\Phi_{770}>$ is
chosen to be singlet under $SU(5)$ $\otimes$ $U(1)_X$ and
orthogonal to 
$<\Phi_{210}>$ in (\ref{vev210_1a}). It is: 
\begin{equation}
<\Phi_{770}>=v_{770}\frac{1}
{3\sqrt{5}}~diag(16,\underbrace{-2,\ldots,-2}_{10
~entries},\underbrace{-2,\ldots,-2}_{10
~entries},\underbrace{1,\ldots,1}_{24 ~entries})\equiv v_{210} <770>_{45}.
\label{vev770_1}
\end{equation}

\subsubsection{SO(10) $\rightarrow$ SU(3)$_c \otimes$ SU(2)$_L
\otimes$ U(1)$_Y \otimes$ U(1)$_X$}\label{ss:10_2}
The case we consider in this subsection is a typical example of
several symmetry breaking chains (see the $E(6)$ cases below)
where the {\em vev}s can be easily written down using the {\em
vev}s for  GUT groups which are themselves subgroups of the one under
consideration.  Here we exploit the findings of sec. \ref{a:su5}
on $SU(5)$ symmetry breaking to obtain the required results providing
enough details. In subsequent subsections we simple
write down the results since the method is the same.
  
To accomplish the desired symmetry breaking the {\em vev} has to
be assigned to a component of $\Phi_D$ which is not only a
non-singlet under $SO(10)$ but also under its subgroup $SU(5)$.
In fact, from the discussions in sec. \ref{a:su5} it must
transform as a 24, 75, or 200 of $SU(5)$.

The 54-dimensional $SO(10)$ representation contains an
$SU(5)\otimes U(1)_X$ (24,0) which is appropriate for the
symmetry breaking under consideration.  Under $SU(5)\otimes
U(1)_X$ 10 = (5,2) + ($\overline{5}$,-2). Using (\ref{vev24_1})
one finds
\begin{eqnarray}
<\Phi_{54,24}> &=& v'_{54}\frac{1}{\sqrt{60}}~diag(3,3,-2,-2,-2,3,3,-2,-2,-2)
\nonumber \\ &=& v'_{54} ~diag(<24>_5, <24>_5)
\equiv v'_{54} <54,24>_{10}.
\label{vev54_2}
\end{eqnarray}
Under $SU(5)\otimes U(1)_X$ 210 $\supset$ (24,0) + (75,0).
Bearing in mind 16 = (1,-5) + ($\bar{5}$,3) + (10,-1) and
employing 
(\ref{vev24_1}) and (\ref{vev24_2})
\begin{eqnarray}
<\Phi_{210,24}> &=&
v'_{210}\frac{1}{\sqrt{60}}~diag(0,3,3,-2,-2,-2,6,-4,-4,-4,
\underbrace{1, \ldots, 1}_{6 ~entries}) \nonumber \\   &=& v'_{210}(0, <24>_5, <24>_{10})
\equiv v'_{210} <210,24>_{16}.
\label{vev210_2a}
\end{eqnarray}
Ensuring orthogonality and using (\ref{vev75_1}) one has:  
\begin{eqnarray}
<\Phi_{210,75}> &=&
v''_{210}\frac{1}{3}~diag(0, \underbrace{0,\ldots, 0}_{5
~entries}, 3, 1,1,1,\underbrace{-1, \ldots, -1}_{6 ~entries})
\nonumber \\   &=& v'_{210}\frac{2}{\sqrt 3} (0,\underbrace{0, \ldots,
0}_{5 ~entries}, <75>_{10})\equiv v'_{210} <210,75>_{16}.
\label{vev210_2b}
\end{eqnarray}

The 770 representation of $SO(10)$ contains within it (24,0),
(75,0), and (200,0) submultiplets under  $SU(5)\otimes U(1)_X$.
As already discussed, $<\Phi_{770}>$ can be expressed as
a traceless, diagonal 45$\times$45 matrix. Further 45 = (1,0) +
(10,4) + ($\overline{10}$, -4) + (24,0). 
Using (\ref{vev200_2}) one gets:
\begin{equation}
<\Phi_{770,200}> = v'''_{770}\sqrt{\frac{8}{5}}(0, \underbrace{0,
\ldots , 0}_{10 ~entries},
\underbrace{0, \ldots , 0}_{10 ~entries}, <200>_{24})
\equiv v'''_{770} <770,200>_{45}.
\label{vev770_2c}
\end{equation}

\subsubsection{SO(10) $\rightarrow$ SU(4)$_c \otimes$SU(2)$_L
\otimes$SU(2)$_R$}

Under  SU(4)$_c \otimes$ SU(2)$_L \otimes$ SU(2)$_R $, 10
$\equiv$ (1,2,2) + (6,1,1).  From the tracelessness condition one
can immediately obtain
\begin{equation}
<\Phi_{54}>=v_{54}\frac{1}{\sqrt{60}}~diag(3,3,3,3,
\underbrace{-2, \ldots, -2}_{6 ~entries}).
\label{vev54_3}
\end{equation}

As noted earlier, $<\Phi_{210}>$ can be represented as a
traceless and diagonal 16$\times$16 matrix.   Since 
 16 $\equiv$ (4,2,1) + ($\bar4$,1,2) one can readily identify 
\begin{equation}
<\Phi_{210}>=v_{210}\frac{1}{\sqrt{8}}~diag(
\underbrace{1,\ldots,1}_{8 ~entries}, \underbrace{-1,\ldots,-1}_{8 ~entries}),
\label{vev210_3}
\end{equation}

Similarly, noting  45 $\equiv$ (15,1,1) + (1,3,1) + (1,1,3)+
(6,2,2) under $SU(4)_c \otimes SU(2)_L \otimes SU(2)_R$, one can write
$<\Phi_{770}>$ as: 
\begin{equation}
<\Phi_{770}>=v_{770}\frac{1}{\sqrt{180}}~diag(\underbrace{-4,\ldots,-4}_{15
~entries},\underbrace{-10,\ldots,-10}_{3+3
~entries},\underbrace{5,\ldots,5}_{24
~entries}).
\label{vev770_3}
\end{equation}
The $<\Phi_{54}>$ and $<\Phi_{210}>$
can also be written in a similar 45$\times$45 form 
and care must be taken to ensure that $<\Phi_{770}>$ is orthogonal to them. 

\subsection{E(6)}

The options available for $\Phi_D$ for $E(6)$ GUTs are 650- and
2430-dimensional.

In $E(6)$ $\overline{27} \otimes 27 = 1 \oplus  78 \oplus 650$.
So, $\Phi_{650}$ can be expressed as a 27$\times$27 traceless
diagonal matrix. For this case $Tr(\lambda_{i}\lambda_{j})=3
~\delta_{ij}$. 

Also, $(78 \otimes 78)_{sym} = 1 \oplus  650 \oplus 2430$. Hence
both $<\Phi_{650}>$ and $<\Phi_{2430}>$ can be represented as
78$\times$78 diagonal traceless matrices. For them
$Tr(\lambda_{i}\lambda_{j})=12 ~\delta_{ij}$.

\subsubsection{E(6) $\rightarrow$ SU(2)$\otimes$SU(6)}

Both 650 and 2430 have directions which are singlets under
$SU(2)\otimes SU(6)$.   Under
$SU(2)\otimes SU(6)$ 27 = (2,$\bar6$) + (1,$\overline{15}$).
Therefore one can readily write $<\Phi_{650}>$ for this channel
of symmetry breaking as the 27$\times$27 diagonal traceless
matrix:
\begin{equation}
<\Phi_{650}>=v_{650}\frac{1}{\sqrt{180}}~diag(\underbrace{5,\ldots,5}_{12
~entries},\underbrace{-4,\ldots,-4}_{15
~entries}).
\label{vev650_1}
\end{equation}

The 2430 {\em vev} can be written down using 
78 = (3,1) + (1, 35) + (2, 20) 
and maintaining orthogonality with
$<\Phi_{650}>$ 
one can write 
\begin{equation}
<\Phi_{2430}>=v_{2430}\frac{1}{\sqrt{3640}}~diag(70,70,70,
\underbrace{18,\ldots,18}_{35
~entries},\underbrace{-21,\ldots,-21}_{40
~entries}).
\label{vev2430_1}
\end{equation}

\subsubsection{E(6) $\rightarrow$ SO(10)$\otimes$U(1)$_\eta$}
The 650 representation has a singlet under $SO(10)$ $\otimes$
$U(1)$ which as before can be expressed as a 27$\times$27 matrix.
Under $SO(10) \otimes U(1)$
27 = (1, 4) + (16, 1) + (10, -2). Using this one finds
\begin{equation}
<\Phi_{650}>=v_{650}\frac{1}{12\sqrt{5}}~diag(40,
\underbrace{-5,\ldots,-5}_{16~entries}, \underbrace{4,\ldots,4}_{10 ~entries}).
\label{vev650_2}
\end{equation}

To write down $\Phi_{2430}$ we note that the decomposition under
$SO(10) \otimes U(1)$ is 78 = (1, 0) + (45, 0) + (16, -3) +
($\overline{16}$, 3).  Ensuring the requirements of orthogonality
to $<\Phi_{650}>$
 and tracelessness we have
\begin{equation}
<\Phi_{2430}>=v_{2430}\frac{1}{4\sqrt{78}}~diag(-108,
\underbrace{-4,\ldots,-4}_{45
~entries},\underbrace{9,\ldots,9}_{16~entries},
\underbrace{9,\ldots,9}_{16~entries}).
\label{vev2430_2}
\end{equation}

\subsubsection{E(6) $\rightarrow$ SU(5)$ \otimes$U(1)$_\xi
\otimes$U(1)$_\eta$}

The results for this option of symmetry breaking can be obtained
by referring to those in sec. \ref{ss:10_1} for $SO(10)
\rightarrow SU(5) \otimes U(1)_X$. Here the {\em vev} must be
assigned to a direction which is a singlet under $SU(5) \otimes
U(1)_\xi \otimes U(1)_\eta$ but not under $SO(10) \otimes
U(1)_\eta$. Such possibilities are the following: 650 of $E(6)$
contains the submultiplets (54,0) and (210,0) under the latter
group and the 2430 of $E(6)$ includes (210,0) and (770,0). As
already noted the $SO(10)$ 54 does not have a singlet direction
of $SU(5) \otimes U(1)_X$. So, we need to consider only the other
possibilities.

It is useful to recall the decomposition 27 = (1,4) + (10,-2) +
(16,1) under $SO(10\otimes U(1)_\eta$. Then from (\ref{vev210_1})
we have
\begin{equation}
<\Phi_{650,
210}>=v_{650}\sqrt{\frac{3}{2}}~diag(0,\underbrace{0,\ldots,0}_{10
~entries}, <210>_{16}).
\label{vev650_3}
\end{equation}
$<\Phi_{650}>$ can also be expressed as a 78$\times$78 traceless
diagonal matrix. Here one uses 78 = (1,0) + (45,0) + (16, -3) +
($\overline{16}$,3) under $SO(10\otimes U(1)_\eta$. Then 
using (\ref{vev210_1}) and (\ref{vev210_1a}):
\begin{equation}
<\Phi_{2430,210}>=v'_{650}~diag(0, <210>_{45}, <210>_{16}, <210>_{16} ).
\label{vev650_3a}
\end{equation}
The remaining {\em vev} is $<\Phi_{2430}>$ which can be written
down using (\ref{vev770_1})
\begin{equation}
<\Phi_{2430,
770}>=v_{2430}\sqrt{\frac{3}{2}}~diag(0, <770>_{45},
\underbrace{0,\ldots,0}_{16 ~entries},
\underbrace{0,\ldots,0}_{16 ~entries} ).
\label{vev2430_3}
\end{equation}

\subsubsection{E(6) $\rightarrow$ SU(3)$_c \otimes$SU(2)$_L
\otimes$U(1)$_Y \otimes$ U(1)$_\xi \otimes$U(1)$_\eta$}
For this symmetry breaking we can utilise the results in sec.
\ref{ss:10_2} for $SO(10) \rightarrow SU(3)\otimes SU(2)_L
\otimes U(1)_Y \otimes U(1)_X$.  The relevant submultiplets are
the following: 650 of $E(6)$ contains  (54,0) and (210,0) under
$SO(10) \otimes U(1)_\eta$ and the 2430 of $E(6)$ includes
(54,0), (210,0) and (770,0). The desired symmetry breaking can
occur through the further $SU(5)$ 24, 75, or 200 content of the
$SO(10)$ multiplets, viz., 54 $\supset$ 24; 210 $\supset$ 24 and
75; and 770 $\supset$ 24, 75 and 200.

The explicit forms of the {\em vev}s are not of much use since
ultimately it is the $SU(5)$ representation which fixes the
$\delta_i$ following the results of sec. \ref{s:su5}. So, we refrain
from displaying the {\em vev}s in this case.

\subsubsection{E(6) $\rightarrow$ SU(3)$_c \otimes$SU(3)$_L \otimes$SU(3)$_R$}

As before, $<\Phi_{650}>$ can be written as a 27$\times$27
matrix.  It turns out that the 650 representation has two
directions which are singlet under $SU(3)_c \otimes SU(3)_L
\otimes SU(3)_R \equiv {\mathcal G}_{333}$. Of course, a {\em
vev} in any
one of these directions or linear combinations thereof may be chosen
to break the symmetry. In particular, two linear combinations may
be identified which respect $\delta_{3L} = \pm \delta_{3R}$.
These are of interest from the physics standpoint as they are
respectively even or odd under D-Parity. 

In this option of $E(6)$ symmetry breaking to $SU(3)_c \otimes
SU(3)_L \otimes SU(3)_R$ one has 27 = (1,$\bar3$,3) +
(3,1,$\bar3$) + ($\bar3$,3,1).   The D-even case is:
\begin{equation}
<\Phi_{650}>=v_{650}\frac{1}{\sqrt{18}}~diag(\underbrace{-2,\ldots,-2}_{9
~entries},\underbrace{1,\ldots,1}_{9
~entries},\underbrace{1,\ldots,1}_{9~entries}).
\label{vev650_5e}
\end{equation}
while the D-odd {\em vev} is
\begin{equation}
<\Phi'_{650}>=v'_{650}\frac{1}{\sqrt{6}}~diag(\underbrace{0,\ldots,0}_{9
~entries},\underbrace{1,\ldots,1}_{9
~entries},\underbrace{-1,\ldots,-1}_{9~entries}).
\label{vev650_5o}
\end{equation}

As in the other cases, $<\Phi_{2430}>$ can be written as a
78$\times$78 traceless diagonal matrix. Noting that
78 = (8,1,1) + (1,8,1) + (1,1,8) + (3,3,$\bar3$) +
($\bar3$,$\bar 3$,3) and maintaining orthogonality with
$<\Phi_{650}>$ and $<\Phi'_{650}>$ 
one can write 
\begin{equation}
<\Phi_{2430}>=v_{2430}\frac{1}{\sqrt{234}}~diag(\underbrace{9,\ldots,9}_{8
~entries},\underbrace{9,\ldots,9}_{8
~entries},\underbrace{9,\ldots,9}_{8
~entries},\underbrace{-4,\ldots,-4}_{27
~entries},\underbrace{-4,\ldots,-4}_{27
~entries}).
\label{vev2430_5}
\end{equation}


\begin{thebibliography}{99}

\bibitem{guts}  J. C. Pati and A. Salam, Phys. Rev. {\bf D10}
(1974) 275; H. Georgi and S. L. Glashow, Phys. Rev. Lett. {\bf
32} (1974) 428; G.~G.~Ross, {\em Grand Unified Theories}
(Benjamin/Cummings, Reading, USA, 1984); R.~N.~Mohapatra, {\em
Unification and Supersymmetry. The frontiers of quark - lepton
physics} (Springer, Berlin, Germany, 1986).

\bibitem{rg} H. Georgi, H. R. Quinn and S, Weinberg,  Phys.
Rev. Lett. {\bf 33} (1974) 451; D. R. T. Jones,  Phys. Rev. {\bf D25} (1982)
581.

\bibitem{prev} Q. Shafi and C. Wetterich, Phys. Rev. Lett. {\bf
52} (1984) 875; C. T. Hill, Phys. Lett. {\bf B135} (1984) 47; L. J.
Hall and U. Sarid, Phys. Rev. Lett. {\bf 70} (1993) 2673.

\bibitem{cr1} J. Chakrabortty and A. Raychaudhuri, Phys. Lett.
{\bf B673} (2009) 57. 

\bibitem{cr2} J. Chakrabortty and A. Raychaudhuri,
  Phys.\ Rev.\   {\bf D81} (2010) 055004.

\bibitem{nonug1}   S.~P.~Martin,
  Phys.\ Rev.\  {\bf D79} (2009) 095019.

\bibitem{nonug2} 
  S.~Bhattacharya and J.~Chakrabortty,
  Phys.\ Rev.\  {\bf D81} (2010) 015007.

\bibitem{so10} H. Georgi, in {\em Particles and Fields -- 1974},
ed. C. A. Carlson (AIP, New York, 1975); H. Fritzsch and P.
Minkowski, Ann. Phys. (N.Y.) {\bf 93} (1975) 193; T. Clark, T. Kuo
and N. Nakagawa, Phys. Lett. {\bf B115} (1982) 26; C. S. Aulakh
and R. N. Mohapatra, Phys. Rev. {\bf D28} (1983) 217.

\bibitem{lrs} J. C. Pati and A. Salam, Phys. Rev. {\bf D10} (1974)
275; R. N. Mohapatra and J. C. Pati, Phys. Rev. {\bf D11} (1975)
566; R. N. Mohapatra and J. C. Pati, Phys. Rev. {\bf D11} (1975)
2558; G. Senjanovi\'c and R. N. Mohapatra, Phys. Rev. {\bf D12}
(1975) 1502.

\bibitem{fsu5}
  S.M. Barr,   Phys.\ Lett.\   {\bf B112} (1982) 219;
I.~Antoniadis, J.~R.~Ellis, J.~S.~Hagelin and D.~V.~Nanopoulos,
  Phys.\ Lett.\   {\bf B194} (1987) 231.

\bibitem{dpar} D. Chang, R. N. Mohapatra and M. K. Parida, Phys.
Rev. Lett. {\bf 52} (1984) 1072; Phys. Rev.  {\bf D30} (1984)
1052. 

\bibitem{e6} F. G\"{u}rsey, P. Ramond and P. Sikivie, Phys. Lett.
{\bf B60} (1976) 177; Y. Achiman and B. Stech, Phys. Lett. {\bf
B77} (1978) 389; J.~L.~Hewett, T.~G.~Rizzo and J.~A.~Robinson,
Phys.\ Rev.\   {\bf D33} (1986) 1476; R.~Howl and S.~F.~King,
JHEP {\bf 0801} (2008) 030.

\bibitem{su6}
  J.~E.~Kim, Phys.\ Lett.\   {\bf B107} (1981) 69; A.~Sen, Phys.\
Rev.\   {\bf D31} (1985) 900; Z.~Chacko and R.~N.~Mohapatra,
Phys.\ Lett.\   {\bf B442} (1998) 199.

\bibitem{e6_333} An early work in this direction is U. Sarkar and
A. Raychaudhuri, preprint CUPP/82-5 (unpublished). 




\end{thebibliography}
\end{document}